\documentclass[superscriptaddress,twocolumn, prb,longbibliography]{revtex4-2}
\usepackage[utf8]{inputenc}
\usepackage{bm}
\usepackage{graphicx}
\usepackage{color}
\usepackage{amsmath}
\usepackage{dsfont}
\usepackage{amsfonts}
\usepackage{amssymb}
\usepackage[colorlinks=true, citecolor=blue, linkcolor=blue, urlcolor=blue]{hyperref}
\usepackage{orcidlink}
\usepackage{etoolbox}
\usepackage{cases}
\usepackage{appendix}
\usepackage{placeins}
\usepackage{braket}
\usepackage{bbold}
\newcommand{\Tr}[1]{\mathrm{Tr}\left(#1\right)}
\usepackage{tikz}
\usetikzlibrary{decorations.pathreplacing}

\begin{document}

\title{Random matrix perspective on probabilistic error cancellation}

\author{Leonhard Moske \orcidlink{0009-0006-9285-7248}}
\affiliation{Institute of Physics, University of Bonn, Nu\ss allee 12, 53115 
Bonn}
\author{Pedro Ribeiro}
\affiliation{CeFEMA-LaPMET, Departamento de Fisica, Instituto Superior Tecnico,
Universidade de Lisboa, Av. Rovisco Pais, 1049-001 Lisboa, Portugal}
\author{Toma\v{z} Prosen}
\affiliation{Faculty of Mathematics and Physics, University of Ljubljana, 
Jadranska 19, SI-1000 Ljubljana, Slovenia}
\author{Sergiy Denysov}
\affiliation{Department of Computer Science, OsloMet – Oslo Metropolitan University, Pilestredet 52, Oslo, N-0130, Norway}
\author{Karol \.{Z}yczkowski}
\affiliation{Faculty of Physics, Astronomy and Applied Computer Science, Jagiellonian University, 30-348 Kraków, Poland}
\affiliation{  Center for Theoretical Physics, Polish Academy of Sciences, Al. Lotników 32/46, 02-668 Warszawa, Poland}
\author{David J. Luitz \orcidlink{0000-0003-0099-5696}}
\affiliation{Institute of Physics, University of Bonn, Nu\ss allee 12, 53115 Bonn}

\begin{abstract} 
 
Probabilistic error cancellation is an attempt to reverse the effect of 
dissipative noise channels on quantum computers by applying unphysical channels 
after the execution of a quantum algorithm on noisy hardware. We investigate on 
general grounds the properties of such unphysical quantum channels by 
considering a random matrix ensemble modeling noisy quantum algorithms. We show 
that the complex spectra of denoiser channels inherit their structure from random 
Lindbladians. Additional structure imposed by the locality of noise channels of 
the quantum computer emerges in terms of a hierarchy of timescales.

\end{abstract}

\maketitle
 
\section{Introduction} 

Current quantum computing hardware is able to execute quantum circuits on a
large set of qubits, but the target circuit defined by the quantum algorithm is
altered by physical noise channels which prevent the reliable execution of deep
circuits required by applications \cite{Preskill_2018,dinca2025quantumprocesstomographycompressed}. Circumventing this
major obstacle is hence an important issue in the field and is tackled from two
angles: Error mitigation \cite{temme_error_2017,gupta_probabilistic_2023,
berg_probabilistic_2023,
Kandala_2019,Scheiber2025reducedsampling,Ferracin2024efficiently,Takagi2022,Kim2023}
tries to undo or reduce the effect of the noise while fault tolerant error
correction \cite{aharonov_fault_1996,kitaev_quantum_1997,shor_scheme_1995}
introduces redundancy which allows for the correction of errors caused by the
noise. It is likely that a combination of both approaches is required for a
true breakthrough in the near term.

While dissipative processes are ubiquitous in hardware, their precise nature is
unknown and depends on the involved qubits and gates in the circuit. This poses
a major challenge for probabilistic error cancellation and extracting the
dominant processes is an expensive task, although it can be reduced to a
simpler problem by techniques such as Pauli twirling
\cite{berg_probabilistic_2023,dinca2025quantumprocesstomographycompressed}. 

The standard approach for probabilistic error cancellation inserts an
unphysical channel layer at multiple points in the target circuit
\cite{berg_probabilistic_2023}. In \cite{tepaske_compressed_2023} it was
proposed to instead use a single operator at the end of the circuit. The upside
is that it reduces the number of gates, reducing the required runs of the
circuit. This operator, which retrieves the unitary circuit from the noisy
quantum channel is called \emph{denoiser}.

Here we start from a general point of view and assume no prior information on
the nature of the noise channels and circuits. It is then reasonable to
consider an ensemble of circuits, built from unitary operations (modeling the
target circuit) and noise channels generated by random Lindbladians to ensure
that we get a completely positive trace preserving (CPTP) map. For each noisy
sample circuit we can then define a denoiser channel which recovers the target
circuit in order to investigate properties of such denoisers on general
grounds. 

We show that the ensemble of denoisers can effectively be understood in terms
of the ensemble of random Lindblad generators
\cite{denisov_universal_2019,wang_hierarchy_2020,sa_spectral_2020,timm_random_2009,lange_random-matrix_2021},
which feature a universal spectral support. Completely random Lindbladians
represent only a crude model of physical dissipation processes. We therefore
refine our ensemble by restricting noise to local (few-qubit) processes, where
an emergent hierarchy of dissipation timescales was observed
\cite{wang_hierarchy_2020,sommer_many-body_2021}. Here we show that this
structure reappears in the corresponding denoisers.

\section{Model} 

We introduce a simple model for an ensemble of noisy quantum circuits 
$\Lambda_U$ of depth $m$ acting on $L$ qubits. We work directly in the folded 
picture of matricised superoperators acting on vectorized density matrices.
The model has two key ingredients: Unitary operators $\mathcal{U}_i\in 
\mathbb{C}^{4^L}$ represent the action of unitary gates of the target quantum 
algorithm. For simplicity, we construct these matrices as global random 
unitary matrices, drawn from the Haar measure. One can view a matrix 
$\mathcal{U}_i$ as the action of one or  multiple layers of a circuit consisting 
of local gates.

The noise is introduced by noise channels $\mathcal{N}_i \in \mathbb{C}^{4^L}$ 
after each layer $\mathcal{U}_i$ of the circuit, such that the full channel is 
given by
\begin{equation}
\Lambda_U = \mathcal{N}_m \mathcal{U}_m \dots \mathcal{N}_2
\mathcal{U}_2  \mathcal{N}_1 \mathcal{U}_1 .
\end{equation}
This formulation is similar to the model used in \cite{berg_probabilistic_2023},
where each unitary circuit layer is paired with a quantum channel to model a 
noisy quantum device.

The quantum state of the $L$ qubits is encoded as a density matrix $\rho$, which 
we vectorize as $\ket{\rho}\rangle$. The superoperator formalism allows for a 
direct characterization of the full spectrum of the channel $\Lambda_U$
Each pair of $\mathcal{U}_i$ and $\mathcal{N}_i$ is understood as one out of a 
total of $m$ layers of the noisy circuit. 
\begin{equation}
	\Lambda_U \ket{\rho}\rangle = \mathcal{N}_m \mathcal{U}_m \dots 
    \mathcal{N}_2
\mathcal{U}_2  \mathcal{N}_1 \mathcal{U}_1 \ket{\rho}\rangle.
\end{equation}

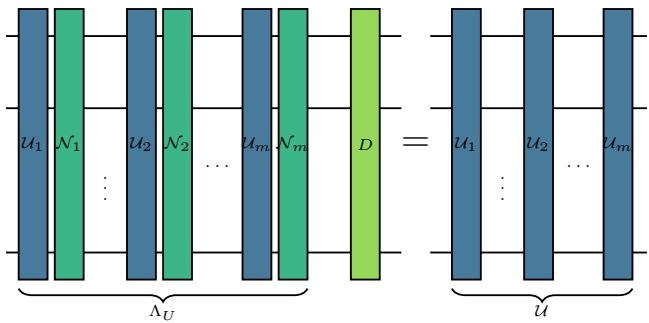
\begin{figure}[tbh!]
\begin{tikzpicture}[scale=1.2, line width=0.7pt]
\definecolor{viridis_g}{HTML}{27AD7D}
\definecolor{viridis_p}{HTML}{89D345}
\definecolor{viridis_b}{HTML}{346B90}
    
\tikzset{font={\fontsize{5.9pt}{12}\selectfont}}
    \def\ysep{0.8}
    \def\xsep{0.40}
    \def\numqubits{4}
    
    \foreach \i in {0,1,3}{
        \draw (0.1,-\i*\ysep) -- (11.2*\xsep,-\i*\ysep);
    }
    
    \node at (3*\xsep,-2*\ysep) {$\vdots$};
    
    \node at (10.2*\xsep,-2*\ysep) {$\vdots$};
    
    \draw[fill=viridis_b!90, draw=black] (0.6*\xsep, 0.3) rectangle (1.4*\xsep, -3*\ysep-0.3);
    \node at (1*\xsep, -1.5*\ysep) {$\mathcal{U}_1$};
    
    \draw[fill=viridis_g!90, draw=black] (1.6*\xsep, 0.3) rectangle (2.4*\xsep, -3*\ysep-0.3);
    \node at (2*\xsep, -1.5*\ysep) {$\mathcal{N}_1$};
    
    \draw[fill=viridis_b!90, draw=black] (3.6*\xsep, 0.3) rectangle (4.4*\xsep, -3*\ysep-0.3);
    \node at (4*\xsep, -1.5*\ysep) {$\mathcal{U}_2$};
    
    \draw[fill=viridis_g!90, draw=black] (4.6*\xsep, 0.3) rectangle (5.4*\xsep, -3*\ysep-0.3);
    \node at (5*\xsep, -1.5*\ysep) {$\mathcal{N}_2$};
    
    \node at (6.15*\xsep, -1.8*\ysep) {$\cdots$};
    
    \draw[fill=viridis_b!90, draw=black] (6.8*\xsep, 0.3) rectangle (7.6*\xsep, -3*\ysep-0.3);
    \node at (7.2*\xsep, -1.5*\ysep) {$\mathcal{U}_m$};
    
    \draw[fill=viridis_g!90, draw=black] (7.8*\xsep, 0.3) rectangle (8.6*\xsep, -3*\ysep-0.3);
    \node at (8.2*\xsep, -1.5*\ysep) {$\mathcal{N}_m$};
    
    \draw[fill=viridis_p!90, draw=black] (9.8*\xsep, 0.3) rectangle (10.6*\xsep, -3*\ysep-0.3);
    \node at (10.2*\xsep, -1.5*\ysep) {$D$};
    
    \draw[decorate,decoration={brace,amplitude=5pt,mirror}] (0.6*\xsep,-3.5*\ysep) -- (8.6*\xsep,-3.5*\ysep) 
        node[midway,below=3pt] {$\Lambda_U$};
    
    \node[font=\Large] at (11.6*\xsep, -1.5*\ysep) {$=$};
    
    \foreach \i in {0,1,3}{
        \draw (12.0*\xsep,-\i*\ysep) -- (18.0*\xsep,-\i*\ysep);
    }
    
    \node at (14.0*\xsep,-2*\ysep) {$\vdots$};
    
    \draw[fill=viridis_b!90, draw=black] (12.6*\xsep, 0.3) rectangle (13.4*\xsep, -3*\ysep-0.3);
    \node at (13.0*\xsep, -1.5*\ysep) {$\mathcal{U}_1$};
    
    \draw[fill=viridis_b!90, draw=black] (14.6*\xsep, 0.3) rectangle (15.4*\xsep, -3*\ysep-0.3);
    \node at (15.0*\xsep, -1.5*\ysep) {$\mathcal{U}_2$};
    
    \node at (16.15*\xsep, -1.8*\ysep) {$\cdots$};
    
    \draw[fill=viridis_b!90, draw=black] (16.8*\xsep, 0.3) rectangle (17.6*\xsep, -3*\ysep-0.3);
    \node at (17.2*\xsep, -1.5*\ysep) {$\mathcal{U}_m$};
    
    \draw[decorate,decoration={brace,amplitude=5pt,mirror}] (12.6*\xsep,-3.5*\ysep) -- (17.6*\xsep,-3.5*\ysep) 
        node[midway,below=3pt] {$\mathcal{U}$};
        
\end{tikzpicture}
\caption{Diagrammatic representation of eq. (\ref{eq:denoiser_def}) defining
the denoiser $D$. Each layer of the noisy circuit $\Lambda_U$ consists of a
unitary operation $\mathcal{U}_i$ followed by a noise channel $\mathcal{N}_i$.
The denoiser $D$ recovers the target unitary circuit $\mathcal{U}$ from the
noisy channel $\Lambda_U$.} \label{fig:denoiser_circuit} 
\end{figure}

\twocolumngrid

We denote the dimension of the Hilbert space of wave functions $N=2^L$.
All superoperators then act on the space of dimension $N^2=4^L$.

For modeling noisy quantum hardware, the regime of \emph{weak} noise is most 
relevant. We therefore model the noise channels $\mathcal{N}_i$ using Lindblad 
generators $\mathcal{L}_i$ acting for short times $t\leq1$. The noise channels 
are then given by:
\begin{equation}\mathcal{N}_i = \exp\left(t\mathcal{L}_i\right).\end{equation}
The timescale $t$ allows us to tune the strength of the noise and $t=0$ 
corresponds to the unitary limit in which we have no noise at all.

Modeling the effect of noise by Markovian dynamics appears to be well justified 
on transmon platforms as confirmed by full process tomography 
\cite{berg_probabilistic_2023,dinca2025quantumprocesstomographycompressed}.

In the folded picture, the unitary operators are given by the tensor product
\begin{equation}\mathcal{U}_i = U_i \otimes U_i^T\end{equation} and we sample
	the factors $U_i$ i.i.d. as random unitary matrices from the Haar
	measure, i.e. from $\mathrm{CUE}(2^L)$. This means that $U_i$ is a
	global operation on the set of qubits and our model features minimal
	structure.

We define the \emph{denoiser} $D$ as the operator that recovers the unitary 
target
quantum circuit $\mathcal{U}$ from $\Lambda_U$: 

\begin{equation}
	\label{eq:denoiser_def}
    D \Lambda_U = \mathcal{U}.
\end{equation}
Here
$\mathcal{U} = \mathcal{U}_m \dots \mathcal{U}_2 \mathcal{U}_1$ is the
combination of the layers without noise, i.e. the target circuit. It can also be 
recovered by setting the dissipation time to $t=0$. We show a diagram of the
equation defining the denoiser in Fig. \ref{fig:denoiser_circuit}.
It is important to note that, since the system undergoes irreversible 
decoherence due to the environment, the denoiser can not be a completely 
positive, trace preserving map and is hence an unphysical channel. The denoiser 
can formally be defined using the inverse of $\Lambda_U$:
\begin{equation}
    D = \mathcal{U} \Lambda_U^{-1}.
\end{equation}

This implies that the denoiser can only be defined for invertible channels 
$\Lambda_U$, i.e. if the spectrum of $\Lambda_U$ lies on the punctured complex 
plane $\mathbb{C}\setminus 0$.

\section{Numerical investigation of the denoiser spectra}
The Lindbladians \cite{denisov_universal_2019, kossakowski_2024_a4b2q-dgk30,Lindblad1976} are generated using  
\begin{equation*}\mathcal{L}=
\sum_{l,k=0}^{N^2-2}K_{k,l}\left[F_l \otimes F^*_k - \frac{1}{2}\left(F_k^\dag F_l
\otimes \mathbb{1} + \mathbb{1}\otimes F_l^T F_k^*\right) \right]. 
\end{equation*}
The $F_l$ form
an orthogonal basis of the operator space of dimension $N$, satisfying $\Tr{F_l
F_k^\dag} = \delta_{l,k}$, while also being traceless $\Tr{F_l} =\nobreak 0$. The complex Kossakowski matrix
$K$ carries the information on the coupling to the environment and is positive
semi-definite. For purely random Lindblad generators we sample $K$ from the
Wishart ensemble. Concretely it is generated as \cite{article_Karol_Sommers} 
\begin{equation}
	K = N\frac{G^\dag
G}{\Tr{G^\dag G}},
\end{equation}
where $G$ is sampled from the square complex Ginibre ensemble,
meaning a matrix of order $N^2 - 1$ with independent Gaussian entries. The normalization of $K$ is
chosen such that the support of the Lindblad spectrum is centered at $-1$. In
\cite{denisov_universal_2019} it was shown that the shape of the spectra
follows an universal behavior, independent of the sampling of $K$. 

In order to be able to sample random Lindbladians efficiently we diagonalize the
Kossakowski matrix $K = Q^\dag D Q$ to reduce the number of terms in the
Lindblad generators. The diagonalized equation reads 
\begin{equation*}\mathcal{L}_D =
\sum_{\mu = 0}^{N^2-2}  D_{\mu, \mu} \left[F_\mu \otimes F_\mu^* -
\frac{1}{2}\left(F_\mu^\dag F_\mu \otimes \mathbb{1} + \mathbb{1}\otimes F_\mu^T
F_\mu^*\right) \right].
\end{equation*}
With $F_\mu = \sum_{n=0}^{N^2-2} Q_{\mu, n}F_n$. The spectra of
the ensemble of Lindblad generators can be found in appendix A, Fig. \ref{fig:Lindbladian}.

The $\mathcal{U}_i$ are uniform random unitary matrices with respect to the
Haar measure. This is done by populating a matrix $H$ with random complex
numbers sampled from a standard normal distribution, performing a QR
decomposition $H=QR$, and then taking the matrix $Q$ as a sample for
$\mathcal{U}_i$, discussed in
\cite{fasi_sampling_2021,mezzadri2007generaterandommatricesclassical}.

\begin{figure}[h!] 
	\includegraphics[width=1.\columnwidth]{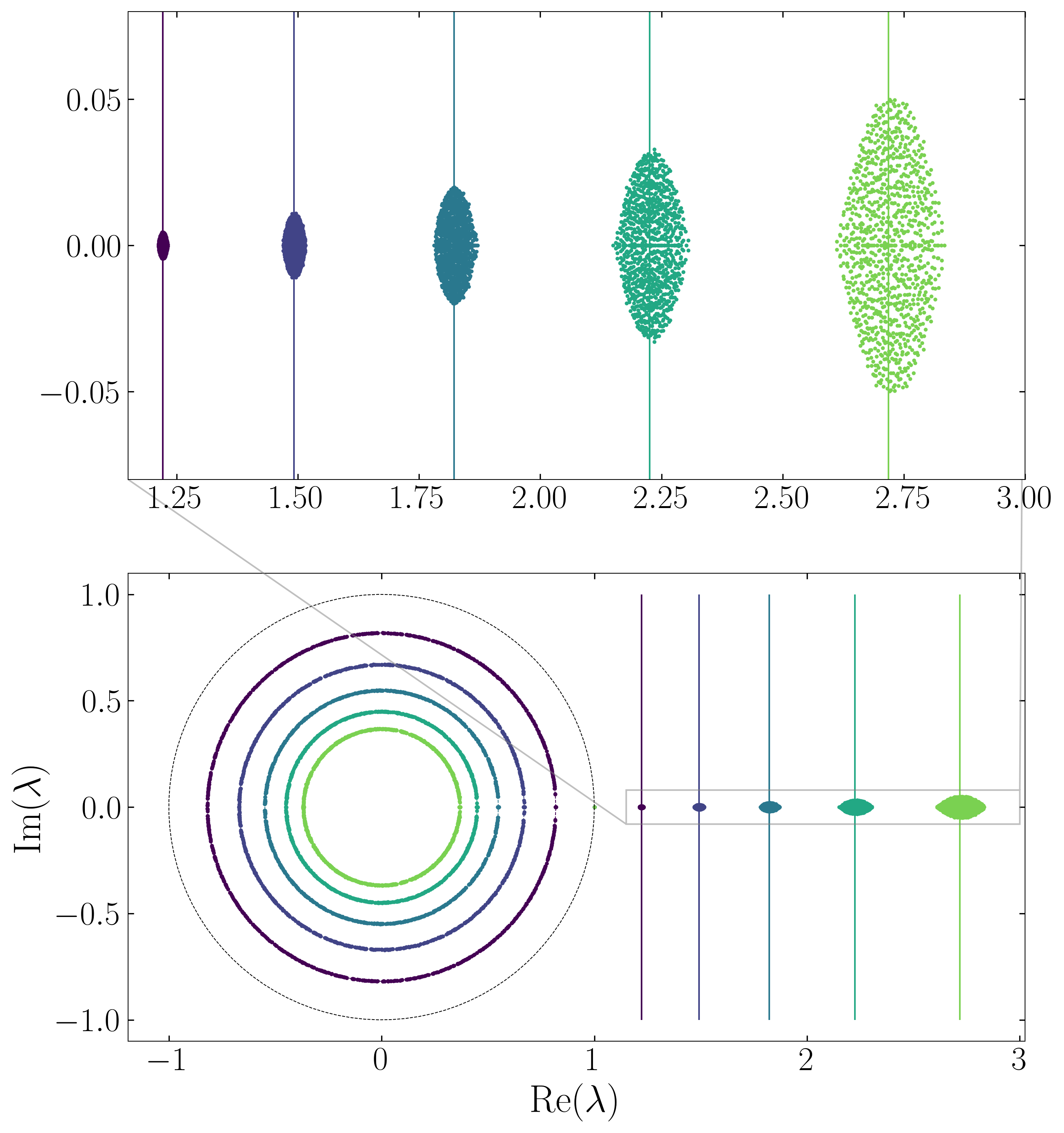}
	\caption{Spectra of $\Lambda_U$ and the denoiser $D$ for different
	noise parameters $t= [0.1, 0.2, 0.3, 0.4, 0.5]$, with $N=32$ and $m=2$.
	The spectra of the noisy channel lie inside the unit circle, while the
	denoiser spectra lie outside, centered around the real axis. For small
	noise $t$ the spectrum of the noisy channel is close to the unit
	circle, while the denoiser is close to the identity with eigenvalue 1.
	With increasing $t$ the noisy channel spectrum moves towards the center
	of the unit circle, while the denoiser spectrum increases in modulus
	and moves away form 1. We draw the predicted centers of the denoiser
	for different $t$ as vertical lines at $\exp(mt)$, according to Eq. (\ref{eq:denoiser_spectrum_prediction}).
	}
\label{fig:denoiser_global_diff_t} \end{figure}

The denoiser is then directly calculated as 
\begin{equation}
	D = \mathcal{U}\Lambda_U^{-1}.
\end{equation}

The spectra of chosen noisy circuits and the corresponding denoisers are shown
in Fig. \ref{fig:denoiser_global_diff_t}. Since for $m=1$ the denoiser only
corresponds to an inverse of $\mathcal{N}_1$, we set the number of circuit layers to $m \geq 2$.

We see that the spectra of the noisy circuit for small $t$ is close to the unit
circle. This is expected as for $t$ going to $0$ the noisy circuit is
approaching the unitary case. The denoiser eigenvalues have modulus greater
than $1$, highlighting the fact that the denoiser is an unphysical operator. As
a consequence of quantum channels being completely positive (i.e. preserving
Hermiticity) the eigenvalues of $\Lambda_U$ come in conjugate pairs
\cite{CHOI1975285}. With $t$ increasing, the eigenvalues of the noisy circuit
decrease in modulus, which implies increasing decay rates with only the steady
state at 1 surviving in the $t \to \infty$ case. In consequence the eigenvalues
of the denoiser increase in modulus with growing $t$.

\begin{figure}[htb!] 
	\includegraphics[width=1.\columnwidth]{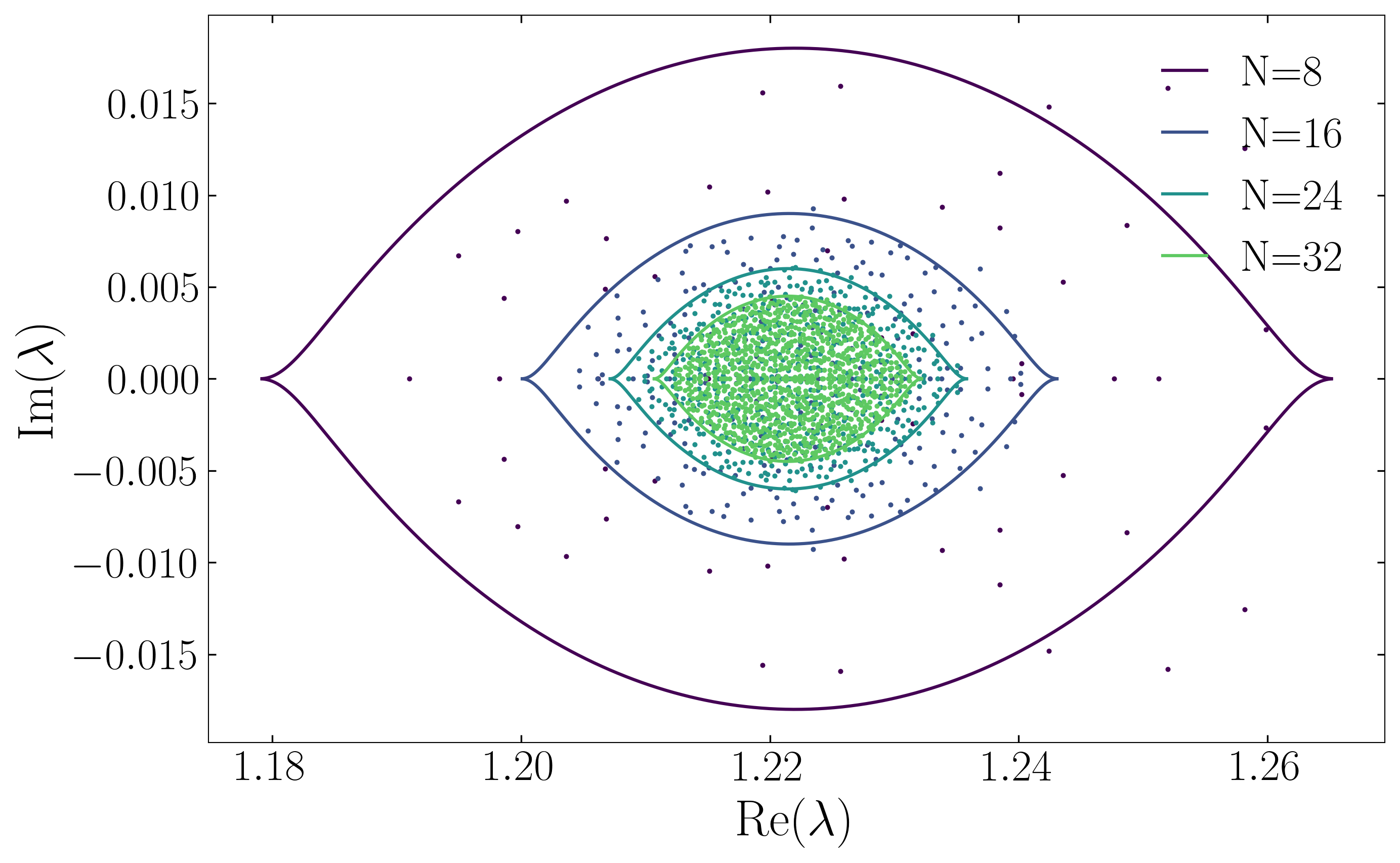}
	\caption{denoiser spectra for different system sizes $N=[8,16,24,32]$. Here $t=0.1$
	and $m = 2$. For readability we exclude the stationary point at 1 and
	we include the prediction of the contour discussed in the next
	section.}
\label{fig:denoiser_diff_N} \end{figure}

The denoiser spectra for different system sizes $N$ is shown in Fig. 
\ref{fig:denoiser_diff_N}. For larger system sizes the support of the spectrum
condenses. This is a result of the behavior of the Lindblad spectrum, which
also condenses around $-1$ with extent $2/N$ as shown in
\cite{denisov_universal_2019}.

\begin{figure}[htb!] 
	\includegraphics[width=1.\columnwidth]{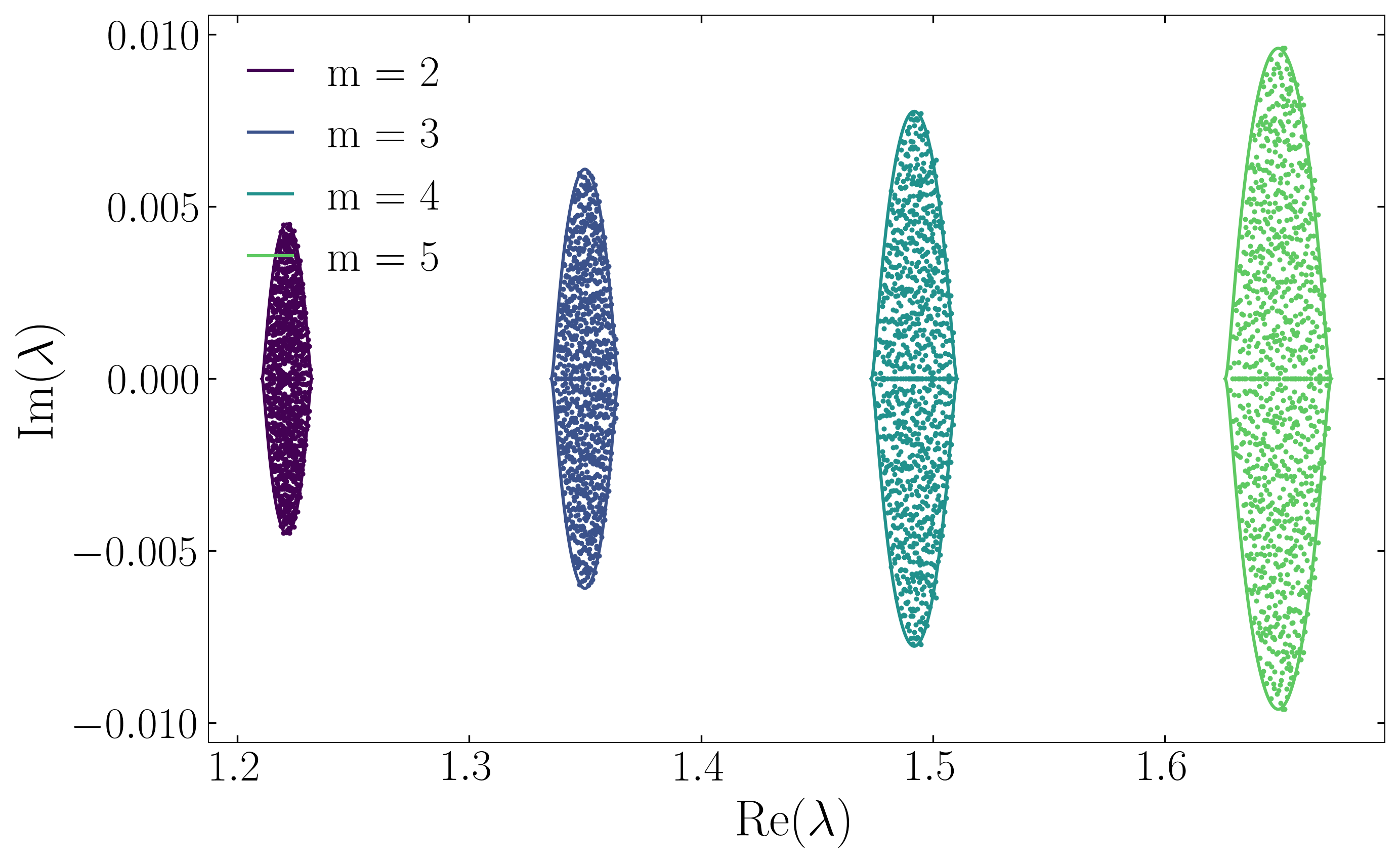}
	\caption{denoiser spectra for different $m=[2,3,4,5]$. Here $t=0.1$ and $N =
	32$. For readability we exclude the stationary point at 1 and we
	include the prediction of the contour discussed in the next section.}
\label{fig:denoiser_diff_N24} \end{figure}

In Fig. \ref{fig:denoiser_diff_N24} we show the denoiser spectra for different
number of layers $m$. For larger $m$ the denoiser eigenvalues increase in
modulus and the support of the spectrum expands. We will explain this behavior
in the next section. Intuitively, more layers of noise need to be inverted by
the denoiser, which leads to larger eigenvalues.

\FloatBarrier
\section{reformulation of denoiser using the Baker-Campbell-Hausdorff formula} 

In the previous section we observed different behaviors of the spectral support
of the denoiser with respect to the parameters of the model $N, t$ and $m$.
To understand the different behaviors we want to
reformulate the noisy channel.

First we recognize that we can change the order
of $\mathcal{U}_i$ and $\mathcal{N}_i$ in layer $i$ by using the fact that
$\mathcal{U}_i$ is an unitary operator. 

\begin{align} 
    \mathcal{U}_j \exp\left( t \mathcal{L}_i\right) &= \mathcal{U}_j \exp\left(t \mathcal{L}_i \right)\mathcal{U}_j^\dag \mathcal{U}_j\\
    &= \exp\left(t\mathcal{U}_j \mathcal{L}_i \mathcal{U}_j^\dag\right) \mathcal{U}_j\\
    &= \exp\left(t\tilde{\mathcal{L}}_i\right)\mathcal{U}_j.    
\end{align}

The only difference is that the Lindbladian is transformed by the unitary, but 
since the unitary is sampled from the Haar measure, the ensemble of $\mathcal{L}$ and
$\tilde{\mathcal{L}}_i$ is the same. 
This means that we can rewrite the noisy circuit as

\begin{align} 
    \Lambda_U &= \dots  \exp\left(t\mathcal{L}_2\right) \mathcal{U}_2 \exp\left(t\mathcal{L}_1\right)\mathcal{U}_1\\
    &=  \left[ \prod_{i}^{m} \exp\left(t\tilde{\mathcal{L}}_i\right)\right]\mathcal{U}.
\end{align}

The rotated Lindbladians are given by $\tilde{\mathcal{L}}_i:=
\left(\prod_{j=0}^{i-1}
\mathcal{U}_{m-j}\right)\mathcal{L}_{i}\left(\prod_{j=i+1}^{m}
\mathcal{U}^{\dag}_{j}\right)$. From equation \ref{eq:denoiser_def} we can identify
the inverse of the denoiser as

\begin{align} 
	D^{-1} = \left[ \prod_{i}^{m} \exp\left(t\tilde{\mathcal{L}}_i\right)\right].
\end{align}

Using the Baker-Campbell-Hausdorff (BCH) formula we can combine the exponentials to

\begin{align*}
	D^{-1} = \exp\left(t\left(\sum^m_{i}\tilde{\mathcal{L}}_{i}\right) + \frac{t^2}{2}\left(\sum_{ j<k}\left[\tilde{\mathcal{L}}_{j}, \tilde{\mathcal{L}}_{k}\right]\right) + \mathcal{O}(t^3) \right),
\end{align*}

where the higher order terms include higher order commutators of the
Lindblad generators. We assume that the noise scale $t$ is small, and
approximate the denoiser only with linear terms in $t$,

\begin{align}
	\label{eq:linear_approx}
		D^{-1} = \exp\left(t\sum^m_{i}\tilde{\mathcal{L}}_{i}\right).
\end{align}

The operator can be inverted by negating the exponent to retrieve an
approximation of $D$.

\begin{figure}[htb!] 
	\includegraphics[width=1.\columnwidth]{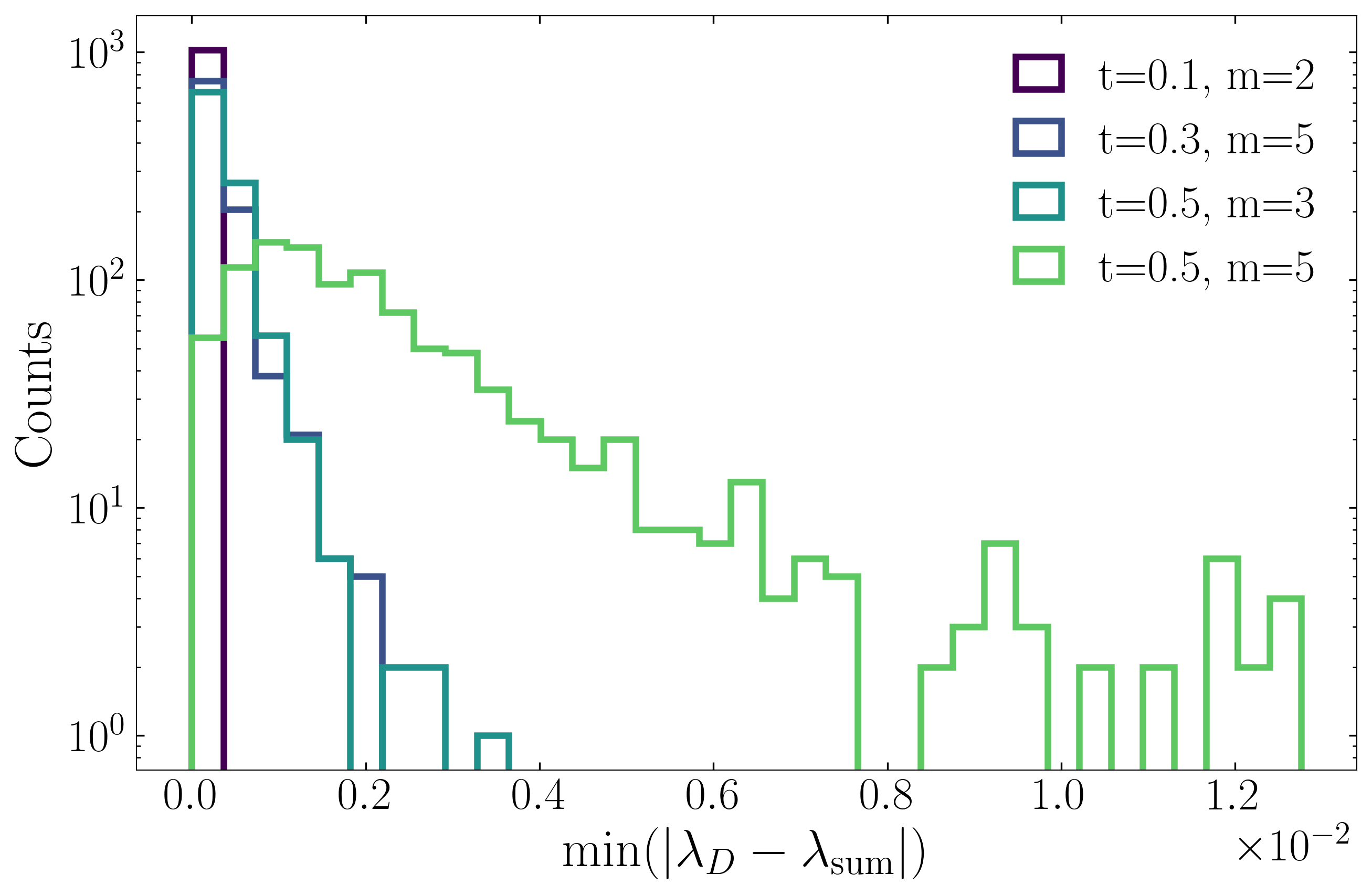}
	\caption{Minimal distance between eigenvalues of the denoiser and the
	spectrum of $\exp\left(-t\sum^m_{i}\tilde{\mathcal{L}}_{i}\right)$, for
	$N=32$ and different combinations of $t$ and $m$ . For each eigenvalue
	we search for the minimal distance to an eigenvalue in the other
	spectrum. For $t=0.1, m=2$ the largest deviation is at the order of
	$10^{-6}$. With $t$ and $m$ increasing the differences of the spectra
	is also increasing.} \label{fig:Comp_denoiser_Sum_Lindbladians_hist}
\end{figure}

In appendix, B Fig. \ref{fig:Comp_denoiser_Sum_Lindbladians} we show the
comparison of the exact denoiser spectra and approximations by the linear term,
for different combinations of $t$ and $m$, showing excellent agreement. Fig.
\ref{fig:Comp_denoiser_Sum_Lindbladians_hist} shows a histogram of the
difference between the eigenvalues of the denoiser and the approximation. We
see that for small $t=0.1$ and small $m=2$ the approximation is very good, with
the differences on the order of $10^{-6}$. For increasing $t$ and $m$ the
eigenvalues start to deviate. With $t$ it is clear that larger values lead to
larger differences, when neglecting higher order terms in the BCH expansion.
For $m$ we note that increasing the number of layers leads to a larger number
of higher order terms in the expansion.

Consequently, to understand the spectrum of the denoiser, we can focus on the
sum of Lindbladians. Here we drop the tilde of the Lindbladians, as we
are not comparing the exact denoiser to approximations, but making an
observation about the ensemble of denoisers. 

First we notice that the sum of
Lindbladians $\sum^m_i \mathcal{L}_i$ is still a Lindbladian:

\begin{align*}
	{\displaystyle
	\sum_{l,n=0}^{N^2-2}\sum^m_i
	\left(K_i\right)_{l,n}\left[F_l \otimes F^*_n - \frac{1}{2}\left(F_n^\dag F_l
	\otimes \mathbb{1} + \mathbb{1}\otimes F_l^T
	F_n^*\right) \right].}
\end{align*}

We thus expect to be able to predict the spectrum of the denoiser with the
contour of the single Lindblad generator. 

The center of the spectral support is
given by the average of the eigenvalues of the Lindbladian, which is
equivalent to the trace of the Lindblad generator divided by the dimension $N^2$. The
trace of the Lindbladian is given by

\begin{align}
    \text{Tr}\mathcal{L} &=  -N\text{Tr}(K) = -N^2.
\end{align}

Since the trace is additive the trace of $m$ Lindbladians is $-mN^2$ and
the average eigenvalue is $\frac{\Tr{\sum^m_i \mathcal{L}_i}}{N^2} = -m$.

Finally, we have to understand the scaling of the support of a sum of
Lindbladians in terms of the number of layers $m$. Here we draw on the results
of \cite{tropp_expected_2015, Sá_2020}, which show that the expected spectral
norm of a sum of independent random matrices scales as $\sqrt{m}$. This means
that the extent of the spectrum of a sum of independent Lindblad generators scales as
$\sqrt{m}$.

With this we can predict the contour of the denoiser spectrum to be similar to
that of a single Lindbladian with the center of the support at $-m$ and
the extent scaling as $\sqrt{m}$.

Given points $\{f_{\mathcal{L},i}\} , i \in \mathbb{N}$ on the contour
of the Lindblad spectrum with center at 0 and extent $2/N$, as described in
\cite{denisov_universal_2019}, the reasoning above shows that the contour of
the denoiser spectrum can be approximated by a map $g$, that connects the
contour of the Lindblad spectrum to the contour of the denoiser spectrum:

\begin{equation}
	f_{D,i} = g \left(f_{\mathcal{L},i}\right) = \exp\left(-t\left(\sqrt{m}f_{\mathcal{L},i} - m\right)\right).
    \label{eq:denoiser_spectrum_prediction}
\end{equation}

\begin{figure}[htb!] 
		\includegraphics[width=1.\columnwidth]{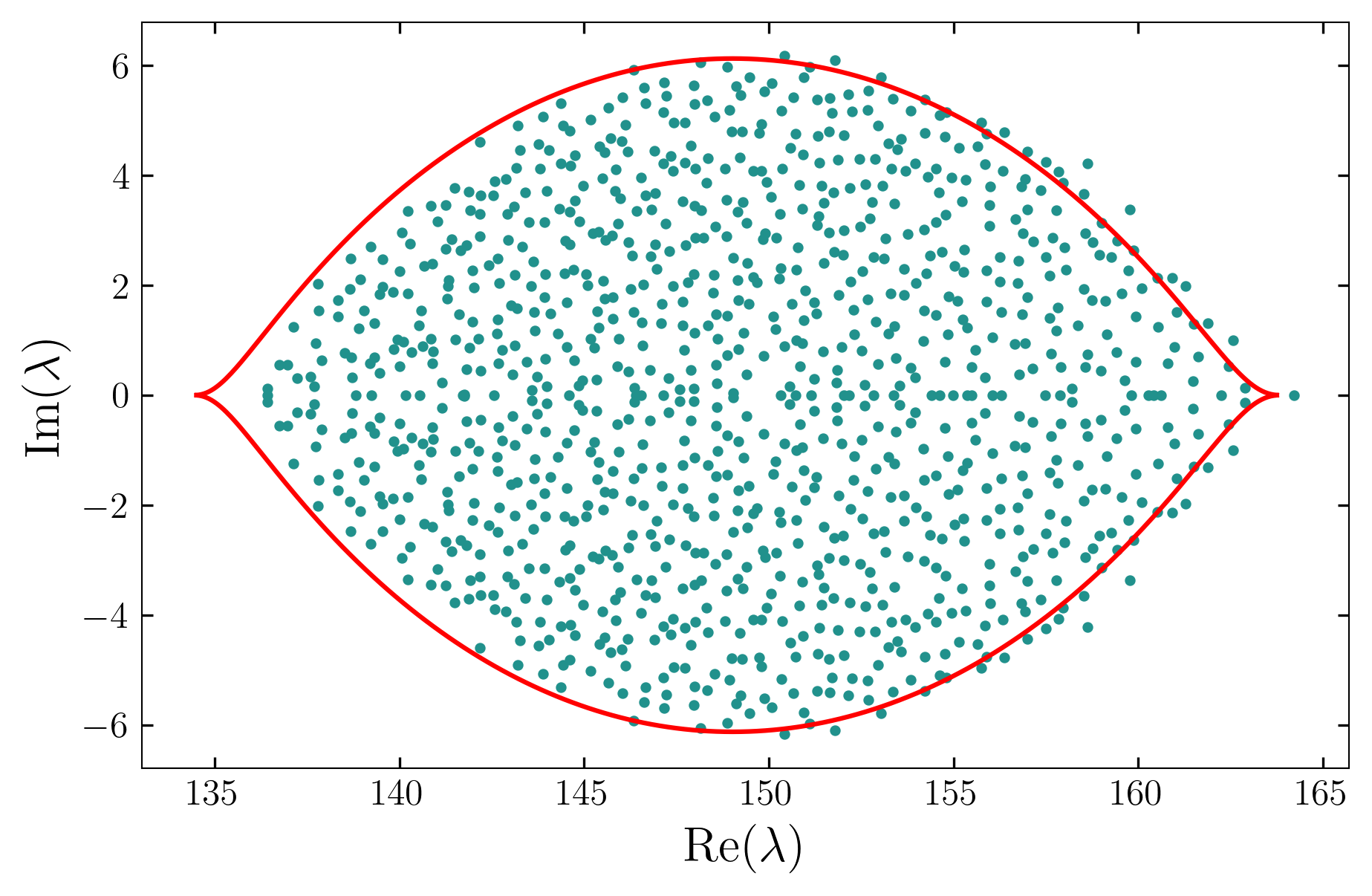}
	\caption{Comparison of the denoiser spectrum and the boundary of its
	support in the complex plane predicted by Eq.
	\ref{eq:denoiser_spectrum_prediction} for $N=32$, $t=0.5$ and $m=10$.
	We see that the denoiser spectrum is well described by the predicted
	contour, which is shown in red. For readability we exclude the steady state at 1.} \label{fig:03_denoiser_spectrum_prediction}
\end{figure}

We can compare this prediction $\{f_{D,i}\}$ with the numerically calculated
spectrum of the denoiser, shown in Fig.
\ref{fig:03_denoiser_spectrum_prediction}. We see that the spectrum of the
denoiser is well described by the prediction and it holds for $t<1$. In
contrast to the comparison of the BCH expansion to the exact denoiser, this
prediction holds for a larger parameter regime, as it only descibes the 
lemon-like shape of the spectral support. 

The only deviation
is to the side of larger modulus, where a few eigenvalues lie outside of the
predicted contour. This is likely due to the finite size effects of the
Lindblad spectrum. We show in appendix A, Fig. \ref{fig:Lindbladian_diff_N} the spectra of
shifted and rescaled Lindblad generators for different $N$. We see that for smaller $N$ the Lindblad
spectrum has eigenvalues outside the contour with smaller modulus. A smaller
modulus in the Lindblad spectrum translates to a larger modulus in the denoiser
spectrum. For increased systemsize $N$ the Lindblad spectrum condenses better
into the predicted contour, which should lead to a better agreement of the
denoiser spectrum with the prediction. 

Additionally the map $g$ implies that the middle of the denoiser spectrum lies
at $\exp(tm)$, which we can confirm in Fig. \ref{fig:denoiser_global_diff_t}.

\section{Local noise} 

The full rank Lindbladian discussed above is an idealized model of the noise.
In other works it was found that the noise in current quantum hardware is best
described by local noise channels \cite{berg_probabilistic_2023,PhysRevResearch.6.033217,Brand2024,PhysRevResearch.5.043210}.
Because of this reason we want to impose a notion of locality on the noise. 

To do so we use Pauli strings $s_l =  \sigma_{x_1} \otimes
\sigma_{x_2} \otimes \dots \otimes \sigma_{x_L}/\sqrt{N} \left(x_i \in
x,y,z,\mathbb{1}\right)$ to form the operator basis $F_l$ of the Lindblad generator.
The number of non-identity Pauli matrices, also known as Pauli weight, $k =
\sum_{i}^{L} \left(1 - \delta_{x_i, 1}\right)$ in the string $s_l$ denotes the
locality of the operator. We say a Lindbladian is $k_{\mathrm{max}}$-local if
all $s_l$ with Pauli weight $k < k_{\mathrm{max}}$ are included in the operator
basis. Consequently, the Kossakowski matrix $K$ has dimension $N_L =
\sum_{k_i}^{k_{\mathrm{max}}} {\binom{L}{k_i}} 3^{k_i}$. We generate $K$ from a
diagonal matrix with positive entries, which is then rotated by a random
unitary matrix, with respect to the Haar measure, on $U(N_L)$. This
ensures that $K$ is positive semi-definite.

\begin{figure}[htb!] 
		\includegraphics[width=1.\columnwidth]{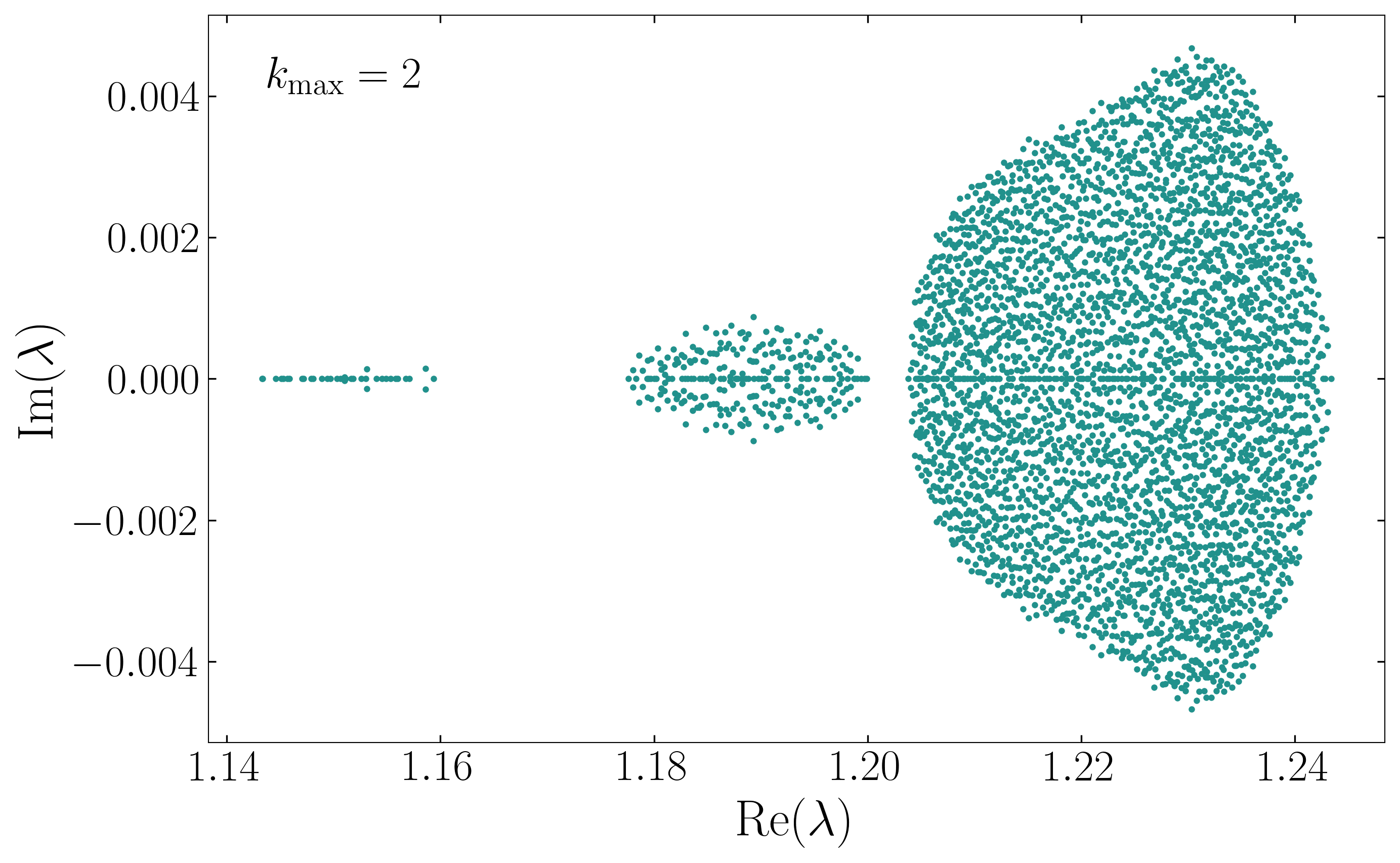}
		\caption{Denoiser spectrum for local Lindblad generators, with $N=64$,
		$t=0.1$, $m=2$ and $k_{\mathrm{max}} = 2$. The spectrum shows multiple
		timescales of decay, which originate from the locality of the
		Lindbladian. } 
		\label{fig:04_local_denoiser}
\end{figure}

The local Lindbladian is then given as,
$$\mathcal{L}_L=
\sum_{l,m=1}^{N_L}K_{m,l}\left[s_l \otimes s^*_m - \frac{1}{2}\left(s_m^\dag s_l
\otimes \mathbb{1} + \mathbb{1}\otimes s_l^T s_m^*\right) \right]. $$

This formulation is equivalent to the
one used in \cite{wang_hierarchy_2020}, where is was found that the spectrum of
these local Lindblad generators show multiple levels of decay depending on the locality
$k_{\mathrm{max}}$. In appendix A Fig. \ref{fig:Lindbladian} we show that we 
reproduce the spectra obtained in \cite{wang_hierarchy_2020}.

\begin{figure}[h!]
	\includegraphics[width=1.\columnwidth]{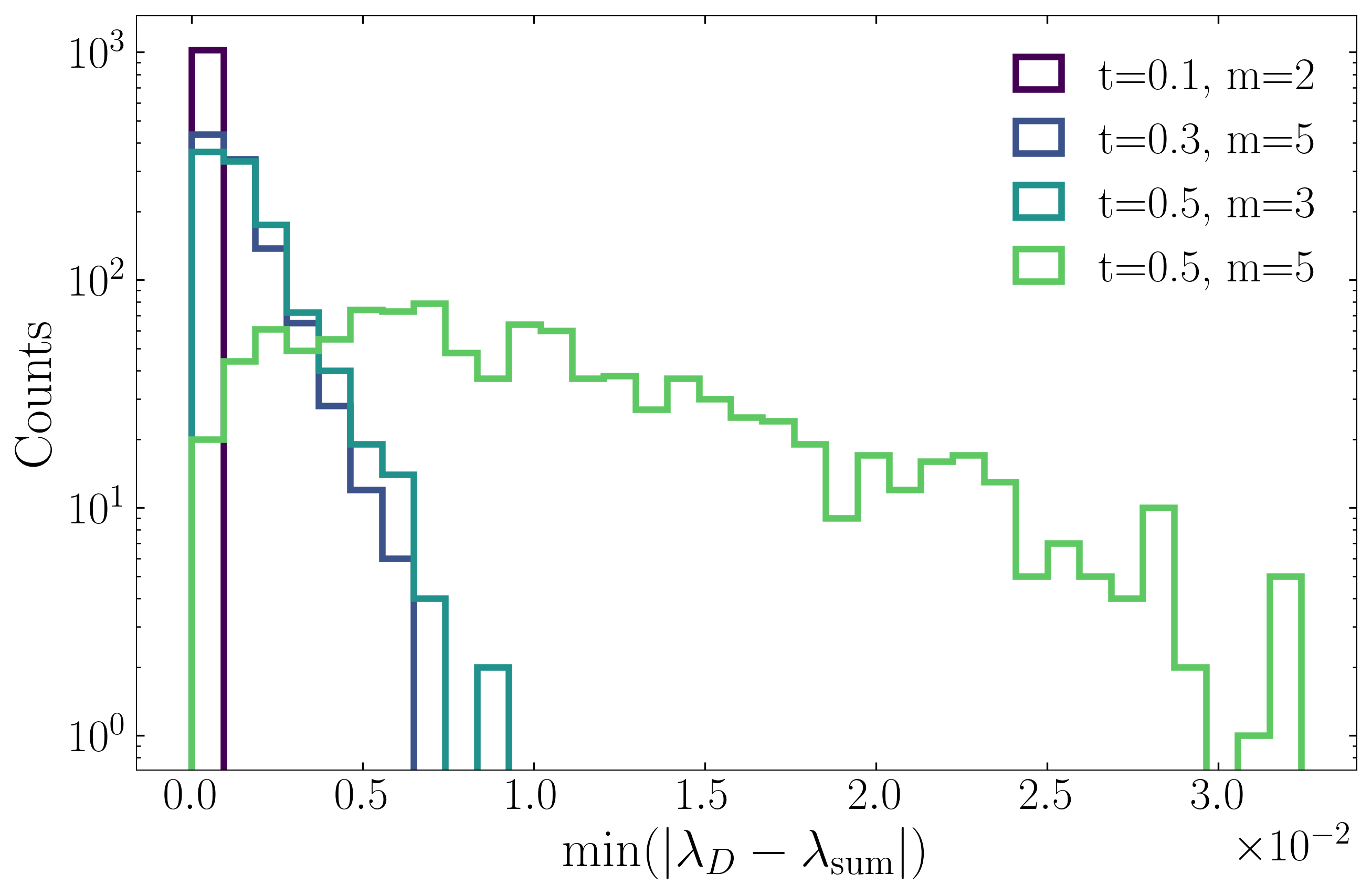}
	\caption{Minimal distance between eigenvalues of the denoiser and the
	spectrum of $\exp\left(-t\sum^m_{i}\tilde{\mathcal{L}_L}_{i}\right)$, for
	$N=32$ and different combinations of $t$ and $m$ . For each eigenvalue
	we search for the minimal distance to an eigenvalue in the other
	spectrum. For $t=0.1, m=2$ the largest deviation is at the order of
	$10^{-4}$. With $t$ and $m$ increasing the differences of the spectra
	is also increasing.}
\label{fig:Comp_denoiser_Sum_Lindbladians_local_hist} \end{figure} 

In order to have a fair comparison between the local noise case and the global
case we normalize the Kossakowski matrix to $N$.
This ensures that the average of the eigenvalues is still at $-1$.

The denoiser spectrum for Lindbladians with locality $k_{\mathrm{max}}=2$ is
shown in Fig. \ref{fig:04_local_denoiser} and we show the other localities in
the appendix E, Fig. \ref{fig:denoiser_local_kmax}. We see that the denoiser
spectrum shows multiple levels in its spectrum. This is a direct consequence of
the multiple levels in the Lindblad spectrum, which are inherited by the
denoiser in accordance with Eq. \ref{eq:linear_approx}. In appendix C we
show the comparison of the denoiser spectrum with the prediction made in
Eq. \ref{eq:linear_approx} in Fig.
\ref{fig:Comp_denoiser_Sum_Lindbladians}. Here we show the histogram of the
difference between the approximation and the exact solution in Fig.
\ref{fig:Comp_denoiser_Sum_Lindbladians_local_hist}. We see that the
approximation is working well also in the local case. 

\FloatBarrier

\section{conclusion} 
We considered an ensemble of random noisy quantum circuits without structure and 
an accompanying denoising operator, which inverts the noise of the environment. 
In the case that the Lindbladian noise is global we showed that the spectrum of 
the denoiser in the complex plane can be predicted by the universal spectrum of a single
random Lindblad generator first shown in \cite{denisov_universal_2019}. 

In the case of few-qubit noise channels, which we consider more realistic for 
current hardware with few-qubit gates, the spectrum of the denoiser exhibits 
different decay rates inherited from the spectra of random few-body Liouvillians 
\cite{wang_hierarchy_2020}.

For a small noise parameter $t$ and for small numbers of layers $m$, the 
denoiser can be expanded to first order in $t$ as an exponential of
the sum of the generating Lindbladians, transformed by the unitary layers of
the circuit, allowing for an analytical connection between the spectra of random 
Liouvillians and denoisers to understand the origin of the spectral features.

While denoisers must correspond to unphysical processes and cannot even be
defined for a non-invertible noise channels, our results reveal an unexpected
feature: The locality structure of the underlying noise survives the scrambling
of random circuits. This persistence of locality-induced spectral structure
indicates that, in realistic settings, effective denoisers may be implementable
by relatively shallow circuits built from few-body gates.

\section{acknowledgements}

This project has received funding from the European Union’s Horizon 2020
research and innovation programme under Grant Agreement No. 101017733 and from
the Deutsche Forschungsgemeinschaft (DFG) through the project DQUANT
(project-id 499347025). We acknowledge support from the DFG through the
Collaborative Research Centers CRC TR185 OSCAR (No. 277625399) and  CRC 1639 
NuMeriQS (project-id 511713970) as well as through the cluster of excellence ML4Q
(EXC 2004, project-id 39053476. 

K{\.Z} acknowledges support by the European Union 
under ERC Advanced Grant Tatypic, Project No. 101142236.

The authors gratefully acknowledge the granted access to the Marvin cluster
hosted by the University of Bonn.

\FloatBarrier

\bibliography{citations}

\clearpage
\onecolumngrid
\begin{appendices}

\section{Appendix A: Spectra of Lindbladians}

Here we show the spectra of the Lindblad generators used in the main text.

	\begin{figure}[htb!]

		\includegraphics[width=1.00\columnwidth]{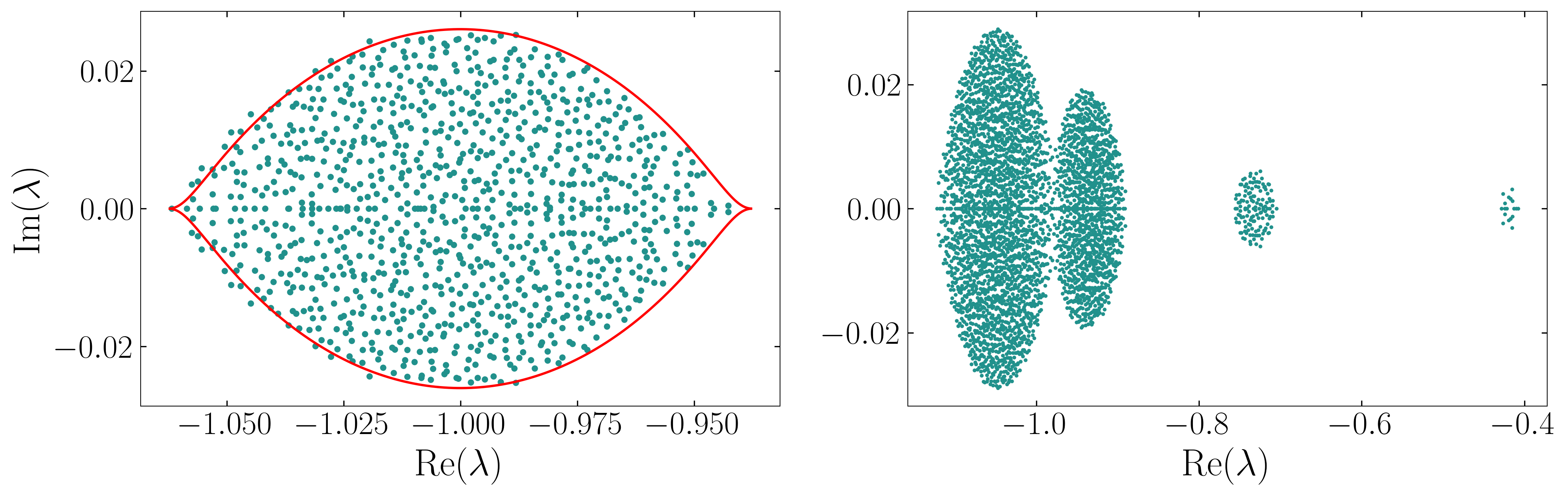}
		\caption{Left: Spectra of Lindbladians, for $N=32$. The
			spectra are resembling the universal behavior with the
			support bounded to the subset of the complex plane
			described in \cite{denisov_universal_2019} and
		represented by a red curve.
		Right: Spectra of local Lindbladians, for $N=64$ and
			$k_{\mathrm{max}} = 2$. We see multiple distinguishable decay
			rates and it reproduces the spectra found in
			\cite{wang_hierarchy_2020}.}
	\label{fig:Lindbladian}
	\end{figure}
\begin{figure}[htb!]
		\includegraphics[width=0.65\columnwidth]{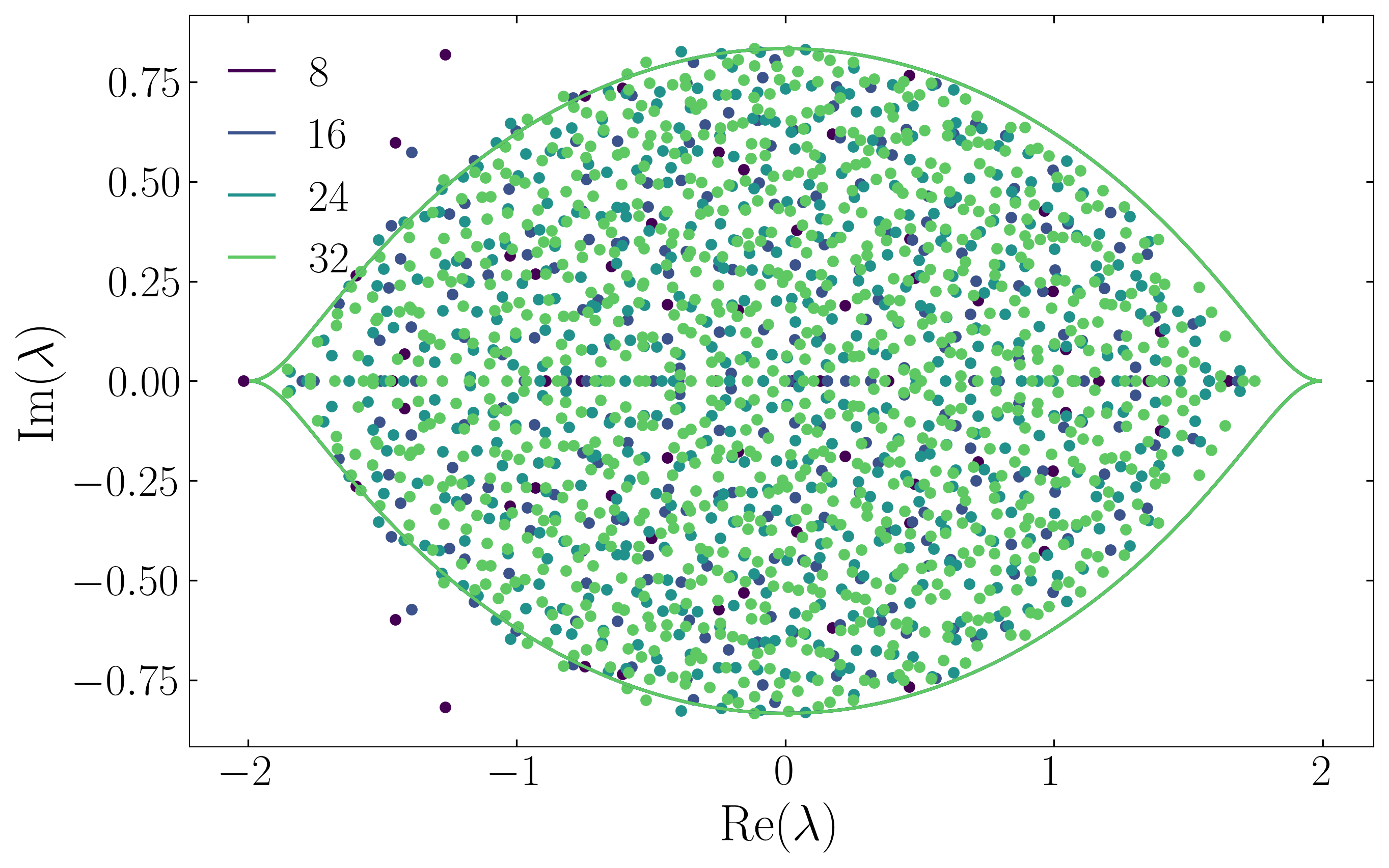}
		\caption{Rescaled and shifted spectra of Lindbladians, for different
		system sizes $N=[8,16,24,32]$. We see that for larger system
		sizes the lemon-like shape derived in
		\cite{denisov_universal_2019} and the spectra agree better,
		indicating a finite size effect.}
		\label{fig:Lindbladian_diff_N} 
	\end{figure}

In Fig. \ref{fig:Lindbladian_diff_N} we show the spectra of global Lindblad generators
		for different system sizes. For smaller system sizes
		we see deviations from the random matrix theory. Since the
		spectra agree better for larger system sizes we conclude that
		this is due to the finite size effect. 
\section{Appendix B: Trace of Lindbladians}

Here we show the short calculation of the trace of a Lindblad generator.

\begin{align*} 
	\text{Tr}\mathcal{L} &= \text{Tr}\left( \sum_{l,p=0}^{N^2-2} K_{p,l}\left[F_l \otimes F^*_p - \frac{1}{2}\left(F_p^\dag F_l \otimes \mathbb{1} + \mathbb{1}\otimes F_l^T F_p^*\right) \right]\right)\\
	&=  \sum_{l,p=0}^{N^2-2} K_{p,l}\left[ \text{Tr}\left(F_l\right)\text{Tr}\left( F^*_p\right) - \frac{1}{2}\left(\text{Tr}\left(F_p^\dag F_l\right)\text{Tr}\left(\mathbb{1}\right) + \text{Tr}\left(\mathbb{1}\right) \text{Tr}\left(F_l^T F_p^*\right)\right) \right]\\
	&=  \sum_{l,p=0}^{N^2-2} K_{p,l}\left[ 0 - \frac{1}{2}\left(\delta_{p,l}N + N\delta_{p,l}\right) \right]\\
&= -N\text{Tr}\left(K\right) \\
	&= -N^2,
\end{align*}

where we used that $\text{Tr}(\mathbb{1}) = N$, $\Tr{F_i^\dag F_j} = \delta_{ij}$ and that the $F_l$ are
traceless. The trace of the Kossakowski matrix is chosen to be $N$, so that the
trace of the sum of Lindbladians is
\begin{align*}
	\text{Tr}\left(\sum^m_{i} \mathcal{L}_i\right) = -N \text{Tr}\left(\sum^m_i K_i\right) = -N^2m. 
\end{align*}
Since the dimension of the Lindblad generator is $N^2$ the center of the spectral
support is at $\frac{\Tr{\sum^m_i \mathcal{L}_i}}{N^2} = -m$. 

\section{Appendix C: Approximating denoiser as sum of Lindbladians}

	\begin{figure}[hbt!]
		\centering
	\includegraphics[width=0.82\columnwidth]{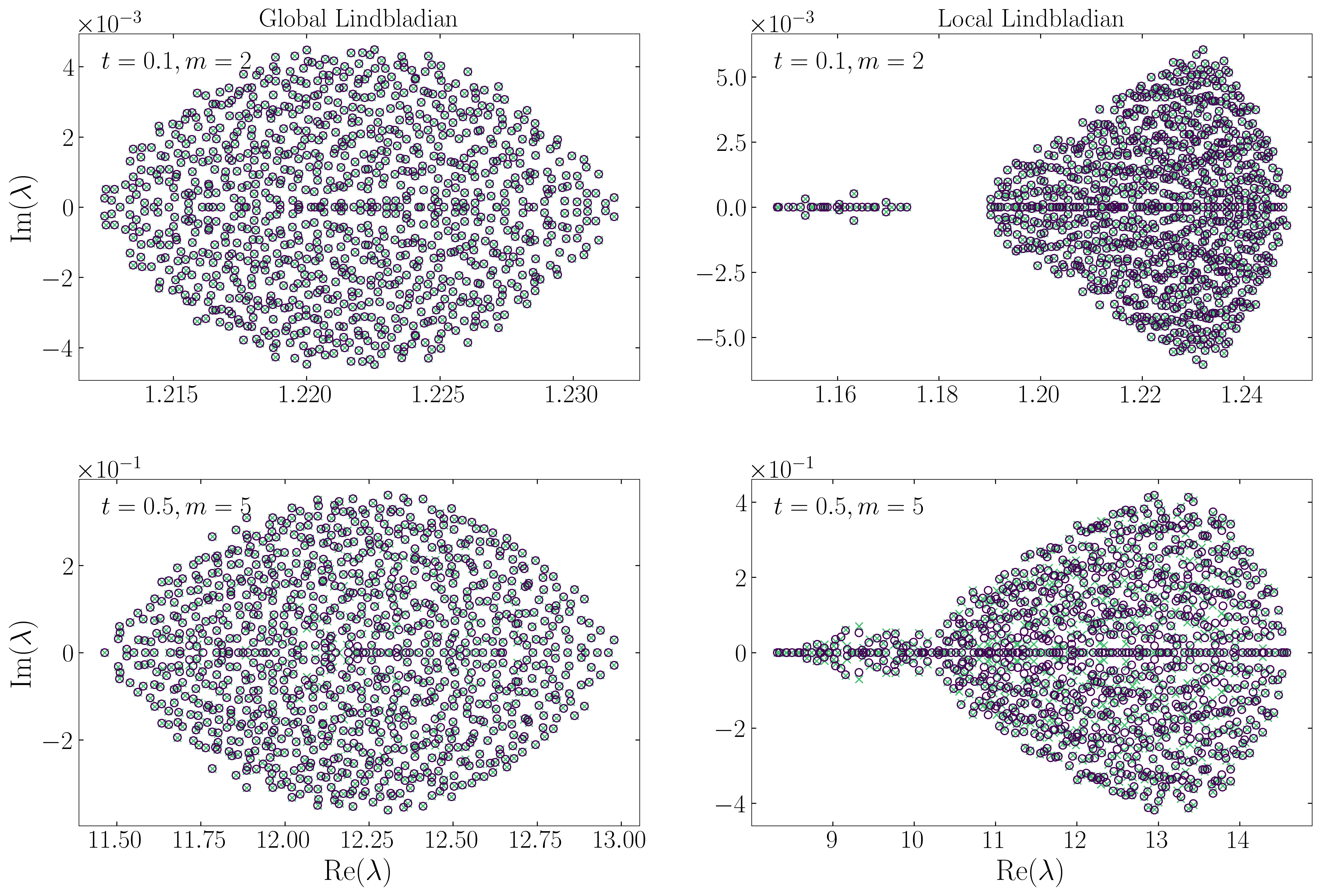}
		\caption{Comparison of the denoiser spectrum (x) and the spectrum of
		$\exp\left(-t\sum^m_{i}\tilde{\mathcal{L}}_{i}\right)$ (o) in the complex plane; for $N=32$, $t=0.1, m=2$
			and $t=0.5,m=5$, for global Lindbladians (left) and local Lindbladian $k_{\mathrm{max} = 2}$ (right). The spectra agree well, showing that the sum of Lindbladians
	is a good approximation of the denoiser for this parameter regime. For larger $t$ and $m$ the eigenvalues start to deviate. For
	readability we exclude the stationary point at 1.}
		\label{fig:Comp_denoiser_Sum_Lindbladians}
	\end{figure}

The spectra of the exact denoiser and the approximations by the sum of
		Lindblad generators (eq \ref{eq:linear_approx}) are shown in Fig.
		\ref{fig:Comp_denoiser_Sum_Lindbladians} for the global and
		local case. We can see that the spectra agree well in both
		cases, showing that the sum of Lindbladians is a good
		approximation of the denoiser for this parameter regime.
		It was found that the commutators of random Lindblad generators
		are small. This explains why the linear approximation is good,
		even for larger $t$. This approximation was found to be most
		sensitive to the number of layers $m$.

	\FloatBarrier
\section{Appendix D: Sum of Kossakowski matrices} 
The denoiser operator is inherently linked to a sum of Lindblad generators, as
it is the leading term in the BCH expansion. To understand the spectrum of the
denoiser, we should understand the spectrum of the sum of Lindbladians. A
sum of Lindbladians can be reduced to a sum of $m$ Kossakowski matrices,
$$\sum^m_i \mathcal{L}_i = \sum_{l,n=0}^{N^2-2}(\sum^m_i K_i)_{l,n}[F_l \otimes F^*_n 
- \frac{1}{2}(F_n^\dag F_l \otimes 1 + 1\otimes (F_n^\dag F_l)^T) ],$$
where the $K$ are sampled as explained in the main text. The idea is then that
the spectrum of the sum of Lindbladians is related to the spectrum of the
sum of Kossakowski matrices. We are interested in the scaling of the spectrum
with the number of layers $m$. The spectra of the Kossakowski
matrices are real, positive and follow the Marchenko-Pastur distribution. We
can use the framework of free probability theory \cite{speicher2014freeprobabilityrandommatrices,Nowak_2017,Voiculescu1992FreeRV} to calculate the spectrum of
the sum of these random matrices. This framework was used  in
\cite{denisov_universal_2019} to calculate the spectrum of the Lindblad
operators. We can use the same framework to calculate the bounds of the
spectrum of the sum of Kossakowski matrices. Since the eigenvalues of the
Kossakowski matrices are real and positive, we can use the R-transform to
calculate the spectrum. The R-transform of the sum of Kossakowski matrices is
given by
$$R(z) = \frac{1}{1-z},$$
where $z$ is a complex number. The R-transform of the sum of Kossakowski
matrices is in turn given by the sum of the R-transforms of the individual
Kossakowski matrices
$$R(z) = \frac{m}{1-z}.$$
The spectrum of the sum of Kossakowski matrices can be calculated from the
Cauchy transform $G(z)$, which is related to the R-transform \cite{Lund01011998,speicher2014freeprobabilityrandommatrices} by
$$R(G(z)) + \frac{1}{G(z)} = z.$$
With the explicit form of the R-transform we can calculate the Cauchy transform
\begin{align*}
\frac{m}{1-G(z)} + \frac{1}{G(z)} &= z,\\
mG(z) + 1 - G(z) &= zG(z) - zG(z)^2,\\
zG(z)^2 - G(z)(m-1 - z) + 1 &= 0.    
\end{align*}

The solution of this quadratic equation is given by
$$G(z) = \frac{(z-m+1) \pm \sqrt{(m-1-z)^2 - 4z}}{2z}.$$
The spectrum is then given as 
$$\rho(z) = \frac{1}{\pi} \lim_{\epsilon \to 0^+} \text{Im}[G(z + i\epsilon)].$$
Since the eigenvalues of the sum $z$ are purely real we can ignore the first
term and focus on the square root. The square root is imaginary if the contents
are negative.
Thus we can search for the roots of $\frac{(n-1-z)^2}{4z^2} - \frac{1}{z}$:
\begin{align*}
 \frac{(m-1-z)^2}{4z^2} - \frac{1}{z} = 0,\\
{(m-1-z)^2} - 4z = 0,\\
z^2 - z(2(m-1) + 4) + (m-1)^2 = 0.   
\end{align*}

The solutions are then the boundaries of the spectrum we are looking for:

\begin{align*}
z &= m  + 1 \pm \sqrt{(m+1)^2 - (m-1)^2},\\
z &= m+1 \pm 2\sqrt{m}.
\end{align*}
We find that the scaling of the boundaries of the spectrum is proportional to
$\sqrt{m}$, which is the same scaling found for the sum of Lindbladians.

\section{Appendix E: Denoiser spectra for different localities $k_{\mathrm{max}}$}

Here we show the denoiser spectra for different localities $k_{\mathrm{max}}$ in 
the Lindblad generators. The spectra feature the characteristic clustering known 
from previous work on random local Lindblad generators 
\cite{wang_hierarchy_2020} depending on the locality $k_{\mathrm{max}}$.

\begin{figure}[htb!]
		\centering
        \includegraphics[width=0.8\columnwidth]{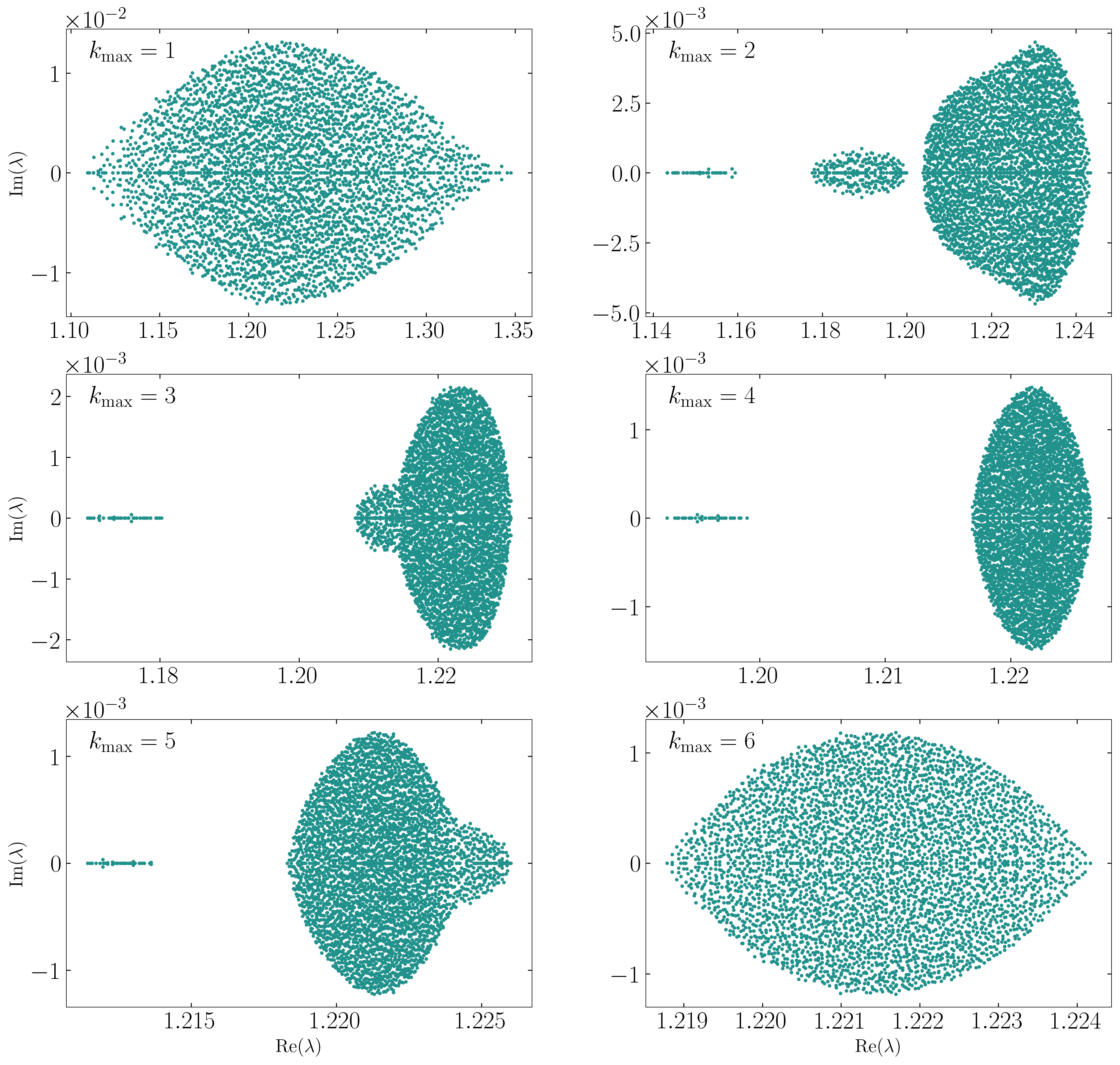}
		\caption{Denoiser spectra for different localities $k_{\mathrm{max}}$ of
		the Lindbladians, with $N=64 = 2^6$, $t=0.1$ and $m=2$. The spectrum
		shows multiple levels of decay, which are related to the
		locality of the Lindbladian. For $k_{\mathrm{max}} = L$ the denoiser
		spectrum reproduces the one of global Lindbladians shown in
		Fig. \ref{fig:denoiser_diff_N24}.}	
	\label{fig:denoiser_local_kmax}
\end{figure}

\end{appendices}
\end{document}